\documentclass[a4paper,10pt]{article}

\usepackage{bm}
\usepackage{latexsym}
\usepackage{dcolumn}
\usepackage{amsmath,amsfonts,amssymb}
\usepackage{graphicx,epsfig}
\usepackage{color}

\def\eq#1{{Eq.~(\ref{#1})}}

\def\gl#1#2{{g_{#1#2}}}

\newcommand{\w}[1]{\bm{#1}}
\newcommand{\el}{\w{\ell}}

\newcommand{\LL}{Lanczos-Lovelock}

\newcommand{\Cal}[1]{\ensuremath{\mathcal{#1}}}

\newcommand{\D}{\ensuremath{\nabla}}

\newcommand{\efe}{Einstein's field equations }

\newcommand{\nsle}{DNS equation }
\newcommand{\nse}{Navier-Stokes equation }

\def\gl#1#2{{g_{#1#2}}}

\def\ch#1#2{{\chi^{#1}_{\phantom{#1}#2}}}    
\def\pip#1#2{{\Pi^{#1}_{\phantom{#1}#2}}}

\title{Action principle for the Fluid--Gravity correspondence and emergent gravity}

\author{Sanved Kolekar\footnote{sanved@iucaa.ernet.in} ~ and T.~Padmanabhan\footnote{paddy@iucaa.ernet.in}\\
IUCAA, Pune University Campus, Ganeshkhind,\\
 Pune 411007, INDIA. \\ 
}

\date{\today}
\begin{document}
\maketitle
\begin{abstract}

It has been known for a long time that Einstein's field equations when projected onto a black hole horizon looks very similar to a Navier-Stokes   equation in suitable variables. More recently, it was shown that the projection of Einstein's equation on to any null surface in any spacetime reduces exactly to the  Navier-Stokes form when viewed in the freely falling frame.  We develop an action principle, the extremization of which leads to the above result, in an arbitrary spacetime. The degrees of freedom varied in the action principle are the null vectors in the spacetime and \textit{not} the metric tensor. The same action principle was introduced earlier in the context of emergent gravity paradigm wherein it was shown that the corresponding Lagrangian can be interpreted as the entropy density of spacetime. 
The current analysis strengthens this interpretation and reinforces the idea that field equations in gravity can be thought of as emergent. We also find that the degrees of freedom on the null surface are equivalent to a fluid with equation of state $PA=TS$. We demonstrate that the same relation arises in the context of a spherical shell collapsing to form a horizon.

\end{abstract}

\section{Introduction}

There is increasing recognition in recent years that the field equations of gravity may have the same conceptual status as the equations of fluid mechanics or elasticity and hence gravity could be thought of as an emergent phenomenon just like, say, fluid mechanics.  (For a recent review, see Ref. \cite{paddyaspects}.) This approach has a long history originating from the work of Sakharov \cite{sakharov} and interpreted in many ways by different authors (for a incomplete sample of references, see Ref. \cite{others}). We will use the term `emergent' in the specific and well-defined sense in terms of the equations of motion, rather than in more speculative vein --- like e.g., considering the space and time themselves to be emergent etc. The evidence for such a \textit{specific interpretation} comes from different facts like the possibility of interpreting the field equation in a wide class of theories as thermodynamic relations \cite{thermo}, the nature of action functional in gravitational theories and their thermodynamic interpretation \cite{holoaction}, the possibility of obtaining the field equations from a thermodynamic extremum principle \cite{aseemtp}, application of equipartition ideas to obtain the density of microscopic degrees of freedom \cite{equiv} etc. 

If the field equations of gravity have the same status as the equations of fluid mechanics, then it should be possible to write down Einstein's field equations (and possibly more general class of field equations; but in this paper we shall confine ourselves to Einstein's gravity in $D=4$.) in a form similar to the equations of fluid mechanics.   It was shown by Damour \cite{damour}  decades ago that this is indeed the case in the context of  black hole spacetimes. He showed that the black hole horizon can be interpreted as a dissipative membrane with Einstein's equations projected on to it taking a form very similar (but not identical) to the Navier-Stokes equation in fluid mechanics. (We shall call these equations Damour-Navier-Stokes equation, or DNS equation for short.) This work formed the basis for the development of membrane paradigm by  several authors to describe black hole physics \cite{membrane}.  Last year, one of us (TP), could generalize this result to \textit{any} null surface in \textit{any} spacetime \cite{paddynavier}. It was shown that, when Einstein's equations are projected on to any null surface and the resulting equations are viewed in the freely falling frame, they become \textit{identical} to Navier-Stokes equation (rather than being very similar to Navier-Stokes equation as in the case of DNS equation). Figuratively speaking, this result shows that a spacetime filled with null surfaces can be equivalently thought of as hosting a fluid, (with the fluid variables related to the structure of the null surface at any given event) which satisfies the Navier-Stokes equation in the local inertial frame (for other related work, exploring the connection between gravity and fluid mechanics, see e.g., \cite{eiling}). There is no  a priori reason for such a mathematical equivalence to arise unless gravitational field equations are emergent  from some, as yet unknown, microscopic structure.

Conventionally, however, one obtains the field equations of the theory by extremizing an action functional for variations of the dynamical variables of the theory. In the case of Einstein gravity one usually obtains the field equation by extremizing the Einstein-Hilbert action with respect to the variations of the metric. In the approach taken in all the previous work in this subject, one first obtains the field equations by this procedure and then projects them on to a black hole horizon (in the original work of Damour, Thorne etc.) or on to a generic null surface (in the context of \cite{paddynavier}). This is, however, conceptually not very satisfactory in the emergent paradigm for two reasons. First, it would be nice if  equations of macroscopic dynamics could be obtained  from a \textit{thermodynamic} extremum principle rather than a \textit{field theoretic} action principle. Second, given the fact that final equations are expressed in terms of variables defined using a null surface, it would be appropriate if the same variables are used in the extremum principle rather than the metric. In other words, we would like to provide a thermodynamic extremum principle from which one directly obtain the Navier-Stokes equation rather than first obtaining or assuming Einstein's field equations and then deriving the DNS equation by  a projection to a null surface.

In this paper we shall show that this can indeed be achieved.  The key to this result lies in the earlier work  \cite{aseemtp} in which it was shown that the field equations  for a wide class of gravitational theories can be obtained by a thermodynamic extremum principle based on null vectors in the spacetime. By an adaptation of this method, we can write down a suitable extremum principle and derive the DNS equation as the resulting Euler-Lagrange equations. What is more, we will show that the functional which is extremised has an interesting interpretation in terms of purely thermodynamic variables defined on the null surfaces. 

The plan of the paper is as follows. In the next section, we shall briefly review the derivation of DNS equation in the conventional procedure and then introduce an entropy functional in terms of the null fluid variables. In Section 2.1, we will obtain the DNS equation by extremizing this entropy functional and compare the result with the more conventional approach. In Section 3, we will provide an interpretation of the extremum principle and derive an equation of state for the null fluid which can be stated simply as $G=E$ where $G$ is the Gibbs free energy and $E$ is the energy, which, of course,  is equivalent to the result $PV=TS$ with $V$ interpreted as the volume of a 2-dimensional surface, ie., the area. Section 4 discusses the nature of this equation of state in full detail. Section 5 provides a brief discussion of the results.

\section{Action for Navier-Stokes Equation}\label{sect1}

We will briefly review the notation and define the geometrical quantities that will be required to study the extrinsic geometry of the null surfaces (for a review see \cite{Gourgoulhon:2005ng}). We will begin by introducing the standard $(1+3)$ foliation of the spacetime with the normals $\w n = - N \w d t$  to $\Sigma_t$ where $N$ is the lapse function.  Let $\w s$ be a unit normal to a set of time-like surfaces such that $\w n \w\cdot \w s =0$. 
We can now define two null vector fields by
\begin{equation}
 \el = N ( \w n + \w s); \qquad \w k = (1/2N) (\w n - \w s)
 \label{stdlk}
\end{equation} 
Here $\w k$ is an auxiliary null vector field with $\el \w \cdot \w k = -1$. We can now define a metric $q_{ab}$ on the 2-dimensional surface $\mathcal{S}_t$ orthogonal to the $\w n$ and $\w s$ through the following standard relations. ($\mathcal{S}_t$ is the intersection of the null surface $\mathcal{S}$ with the time constant surface $\Sigma_t$)
\begin{equation}
 q_{ab} =\gl ab + n_a n_b - s_a s_b = \gl ab+ \ell_a k_b + \ell_b k_a; \quad q_{ab} \ell^b =0 = q_{ab} k^b
\end{equation} 
The mixed tensor $q^a_b$ allows us to project quantities onto $\mathcal{S}_t$. We can also define another projector orthogonal to $k^b$ by the definition 
$\pip db = \delta^d_b  + k^d \ell_b$
which has the properties 
\begin{equation}
\pip ab \ell^b = \ell^a ; \quad \pip ab k^b =0; \quad \pip ab \ell_a =0; \quad \pip ab k_a =k_b.                                                                                    \end{equation} 
The Weingarten coefficients can now be introduced as the projection of the covariant derivative $\nabla_d \ell^a$ by the definition \begin{equation}
\ch ab \equiv \pip db \nabla_d \ell^a = \nabla_b \ell^a + \ell_b ( k^d \nabla_d \ell^a)
\label{defchi1}                                                                                                                                                                                                                        \end{equation} 
which has the following properties
\begin{equation}
 \ch ab \ell^b \equiv \kappa \ell^a; \ \chi_{ab} k^b = 0; \ \chi_{ab} \ell^a = 0; \  \chi_{ab} k^a  \equiv - \omega_b= - \ell^j \nabla_j k_b
\label{defomega}
\end{equation} 
where the surface gravity $\kappa$ is defined through the relation $\ell^j \nabla_j \ell_i = \kappa \ell_i$ and $\omega_a $
through the last equality. $\omega_a$ satisfies the relations $\omega_a \ell^a = \kappa$ and $\omega_a k^a =0$.
 We next define $\Theta_{ab}$ by projecting $\chi_{mb}$ to $\mathcal{S}_t$. We get, on using $\ell^m \chi_{mb} =0 $ and $k^m \chi_{mb} = - \omega_b$, the result:
\begin{equation}
 \Theta_{ab} = q^m_a \chi_{mb} = \chi_{ab} + k_a \ell^m \chi_{mb} + \ell_a k^m \chi_{mb} = \chi_{ab} - \ell_a \omega_b
\label{deftheta2}
\end{equation} 
Using \eq{defchi1} we see that 
\begin{equation}
 \Theta_{ab} = \Theta_{ba} = \nabla_b \ell_a + \ell_a k^i \nabla_i \ell_b - \ell_b \omega_a = q^m_a  q^n_b \nabla_m \ell_n
\label{deftheta}
\end{equation} 
This result shows that $\Theta_{ab}$ is a natural projection of the covariant derivative $\nabla_m \ell_n$ onto the surface $\mathcal{S}_t$ and, obviously, $\Theta_{ab}\ell^b=0=\Theta_{ab}k^b$.
The trace of $\Theta_{ab}$,  denoted by $\theta$, is given by
\begin{equation}
 \Theta^a_a = \theta = \nabla_a l^a - \kappa
\label{deftheta1}
\end{equation} 
It is also convenient  to define a similar projection of $\omega_a$ by $\Omega_b \equiv q^a_b \omega_a$. We have
 \begin{equation}
 \Omega_b \equiv q^a_b \omega_a    = - q^a_b (k_m \ch ma ) = \omega_b  - \kappa k_b ( k_m \ell^m) = \omega_b + \kappa k_b    
\label{defOmega}                                                                                                                                                                                                                                                                                                                                                                                                                                                                                                                 \end{equation} 
These results allow  us to express the projection of Einstein's equations onto $\mathcal{S}_t$. 
To do this one begins with the standard relation
$\nabla_m \nabla_a \ell^m - \nabla_a \nabla_m \ell^m = R_{ma} \, \ell^m $ 
and substitute for $\nabla_a \ell^m$ using \eq{deftheta} and for $\nabla_m \ell^m$ using \eq{deftheta1} repeatedly. This leads, after some straightforward algebra, to the relation 
\begin{eqnarray}
  R_{ma} \, \ell^m & = & \nabla_m \Theta^m_{a} + \ell^m \nabla_m \omega_a + (\kappa + \theta) \omega_a  - \nabla_a (\kappa + \theta) - \Theta_{am} k^n \nabla_n \ell^m \nonumber \\
  & & - \left(\omega_m k^n \nabla_n \ell^m + \nabla_m k^n \, \nabla_n \ell^m  + k^n \nabla_m \nabla_n \ell^m \right) \ell_a
 \label{expansion1}
\end{eqnarray}
The \nsle is obtained by contracting \eq{expansion1} with $q^a_b$. We will state here the final expression which is sufficient for our discussion but the reader may refer to \cite{paddynavier} for a  derivation and detailed discussion.
\begin{eqnarray}
R_{mn} \, \ell^m q^n_{a}  &=& q^m_{a}  {\mathrm{\pounds}}_{\el} \Omega_m + \theta \, \Omega_a -  D_a (\kappa + \theta)  +   D_m \Theta^m_{a} \nonumber\\
&=&q^m_{a}  {\mathrm{\pounds}}_{\el} \Omega_m+ \theta \, \Omega_a -  D_a \left( \kappa + \frac{\theta}{2} \right) +   D_m \sigma^m_{a} = T_{mn} \, \ell^m q^n_{a}
\label{rmneqn}
\end{eqnarray} 
where $ D_a$ is the covariant derivative  defined using the metric on $\mathcal{S}_t$ and ${\mathrm{\pounds}}_{\el}$ denotes the Lie derivative along $\el$. We have also separated out the trace of $\Theta_{mn}$ and define $\sigma_{mn} = \Theta_{mn} - (1/2)q_{mn} \theta$. we see that \eq{rmneqn} has the form  of a \nse for a fluid with the convective derivative replaced by the Lie derivative (This can be taken care of by working in the local inertial frames; see \cite{paddynavier}). We also note that the corresponding fluid quantities are (i) momentum density $-\Omega_a/8\pi$ (for a discussion of this quantity, see Appendix[\ref{momentum}]), (ii) pressure $(\kappa/8\pi)$, (iii) shear tensor $\sigma^m_{a}$ (iv) shear viscosity coefficient $\eta=(1/16\pi)$. (v) bulk viscosity coefficient  $\xi=-1/16\pi$ and (vi) an external force $F_a= T_{ma} \ell^m$. The meaning of $\sigma^m_{a}$ as the shear tensor has been discussed extensively in the literature \cite{paddynavier}. (For a discussion regarding the bulk viscosity coefficient  $\xi$ and the fluid/gravity correspondence in the modern perspective see Section 5). Having identified the relation between fluid variables and quantities describing the extrinsic geometry of the null surfaces we now have a dictionary between the two through Eqs.(\ref{defchi1}-\ref{defOmega} ).

The DNS equation of Eq.(\ref{rmneqn}) was obtained by first writing the Einstein's field equation and then projecting them suitably on the horizon. As we said before, it would be conceptually more satisfying to obtain the DNS equation directly from a variational principle starting from a functional written in terms of variables describing the viscous fluid. (Even in the case of usual fluid mechanics, a corresponding variational principle  for the Navier-Stokes equation is not readily available in the literature.) In the next section, we will describe such an extremum principle based on normals to null surfaces and their derivatives.

\subsection{Obtaining the \nsle}

In this section we show that starting from a suitable lagrangian one can directly derive the \nsle that is, Eq.(\ref{rmneqn}) from an action principle without first assuming the Einstein field equations. We now proceed to investigate the required action. Note that it is sufficient to obtain the right hand side of the Eq.(\ref{expansion1}) equal to $T_{mn} \, \ell^m$ as the Euler Lagrange equation, then the \nsle follows from it by projecting along $q^n_{a}$. To begin with, we take clue from earlier works \cite{aseemtp} where it was shown that an entropy functional  $S_{grav}$ can be associated with every null vector in the spacetime  and by demanding $\delta[S_{grav}+S_{matter}]=0$ for all null vectors in the spacetime, where $S_{matter}$ is the relevant  matter entropy, one can obtain the field equations of gravity in all Lanczos-Lovelock models of gravity. The entropy functional ($S_{grav}+S_{matter}$) was defined to be 
\begin{equation}
S[l^a]=-\int_\Cal{V}{d^4x\sqrt{-g}} \left(4P_{ab}^{cd} \D_cl^a\D_dl^b -  T_{ab}l^al^b\right) \,,
\label{ent-func-2}
\end{equation}
where $P^{cd}_{ab} = (1/2)(\delta^c_a \delta^d_b - \delta^d_a \delta^c_b)$ for Einstein gravity. The origin and properties of the tensor $P^{abcd}$ in a general \LL\ theory has been discussed extensively before in the literature and we refer the reader to \cite{aseemtp} and \cite{paddyaspects} for full details. With the help of our dictionary comprising of Eqs.(\ref{defchi1}-\ref{defOmega}) we can write the above entropy density functional completely in terms of the fluid variables. This can be achieved by first substituting for $\nabla_a l_b$ in terms of $\Theta_{ab}$ using \eq{deftheta} and \eq{deftheta1} and then using the orthogonality condition $\Theta_{ab} l^a =0 $ and the relation $\omega_a l^a = \kappa$ to get
\begin{equation}
S[l^a]=-\int_\Cal{V}{d^4x\sqrt{-g}} \left(-\Theta_{ab}\Theta^{ba} - \kappa^2 + (\theta + \kappa)^2 -  T_{ab}l^al^b\right)
\label{actiondef}
\end{equation}
where the lagrangian for gravity now written in terms of fluid variables is
\begin{equation}
{\cal L}_{grav} = -\Theta_{ab}\Theta^{ba} - \kappa^2 + (\theta + \kappa)^2
\label{lagrangiandef}
\end{equation}

We will first demonstrate that extremising the above action w.r.t the variations of the null normals leads to the \nsle directly. To show this, we first write the gravitational lagrangian in a different but equivalent form: 
\begin{eqnarray}
{\cal L}_{grav} &=& -\Theta_{ab}\Theta^{ba} - \kappa^2 + (\theta + \kappa)^2 \nonumber \\
&=& -\left( \Theta_{ab} +  w_a l_b - l_a k^i \nabla_i l_b\right) \left( \Theta^{ba} +  w^b l^a - l^a k^i \nabla_i l^b\right)+ (\theta + \kappa)^2
\end{eqnarray}
where we have used $\Theta_{ab}l^a = 0$, $l^al_a =0$, $l^a \nabla_m l_a =0$ and $\omega_a \ell^a = \kappa$ to factorize the first expression. Using $\delta(\Theta_{ab} +  w_a l_b - l_a k^i \nabla_i l_b) = \delta^i_a \delta^j_b \delta(\nabla_i l_j)$, we find that the Euler Lagrange derivative $E_j$ for the gravitational lagrangian ${\cal L}_{grav}$ is of the form
\begin{eqnarray}
-E_j &\equiv& \nabla_i \left( \frac{\partial {\cal L}_{grav}}{\partial \nabla_i l^j} \right) - \frac{\partial {\cal L}_{grav}}{\partial l^j} \nonumber \\
&=& -2 \nabla_i \left( \Theta_j^{\ i} +  w_j l^i - l_j k^m \nabla_m l^i\right) + 2\nabla_j (\theta + \kappa) \nonumber \\
&=& -2\nabla_i \Theta^i_{j} -2 l^i \nabla_i \omega_j -2 (\kappa + \theta) \omega_j  +2 \nabla_j (\kappa + \theta) +2 \Theta_{jm} k^n \nabla_n l^m \nonumber \\
  && +2 \left(\omega_m k^n \nabla_n l^m + \nabla_m k^n \, \nabla_n l^m  + k^n \nabla_m \nabla_n l^m \right) l_j \label{eulerderivative}
\end{eqnarray}
By comparing the above expression with \eq{expansion1}, we can see that $E_j = 2 R_j^m l_m$. Hence projecting the Euler Lagrange derivative $E_j$ along $q^j_a$ is equivalent to $2R_{mn} \, \ell^m q^n_{a}$. It is then obvious that the Euler Lagrange equation of the full lagrangian (gravity plus matter) will lead to required \nsle of \eq{rmneqn}; $2R_{mn} \, \ell^m q^n_{a} = T_{mn} \, \ell^m q^n_{a}$. We then use the following two algebraic relations
\begin{eqnarray}             
q^n_{a} \nabla_m \Theta^m_{n} - \Theta^m_{a}  k^n \nabla_n \ell_m   &=& D_m \Theta^m_{a} + \Theta^m_{a}  \Omega_m \\
q^n_{a}  {\mathrm{\pounds}}_{\el} \Omega_n    & = & q^n_{a} \ell^m \nabla_m \omega_n + \Theta_a^{m} \Omega_m + \kappa \Omega_a , 
\end{eqnarray}
to bring \eq{eulerderivative} to the form of the \nsle in \eq{rmneqn}.
\begin{eqnarray}
q^m_{a}  {\mathrm{\pounds}}_{\el} \Omega_m+ \theta \, \Omega_a -  D_a \left( \kappa + \frac{\theta}{2} \right) +   D_m \sigma^m_{a} = T_{mn} \, \ell^m q^n_{a}
\label{extracodn}
\end{eqnarray} 
It has been shown \cite{aseemtp} that the same action when expressed in the form of \eq{ent-func-2} also leads to the \efe with an undetermined cosmological constant on extremising w.r.t to the variations of the null normals and demanding that the extremum condition holds for all null vectors. We refer the reader to \cite{aseemtp} for a full derivation. Such an approach leads to the condition that $(R_{ab} - (1/2)T_{ab})l^a l^b = 0$ for all null vectors $l^a$. Using Bianchi identity and the condition $\nabla_a T^{ab} =0 $, one can show that the above condition is equivalent to the \efe
\begin{eqnarray}
 R_{ij} -  \frac{1}{2} g_{ij}R  = \frac{1}{2}( T_{ij} + g_{ij}\Lambda)
\end{eqnarray}
In this approach one demands the extremum to hold for all null vectors which leads to the \efe which are independent of $l^a$ whereas in the case of \nsle we get an equation in terms of $l^a$ for any null surface. Validity of the DNS equation for all null surfaces is then equivalent to the validity of the \efe. Thus we find that starting from the Lagrangian in Eq.(\ref{lagrangiandef}), whenever \nsle holds, the Einstein field equations also hold starting from the same emergent action. This establishes another way of proving the equivalence between the \efe and the \nse through a common action.

\section{Interpretation of the Action}\label{sect2}

We will now show that the action in \eq{actiondef} acquires a thermodynamic interpretation in  terms of a local entropy density and hence extremising the action could be viewed as equivalent to extremising the entropy density of the spacetime in the emergent gravity paradigm.

We begin by considering the spacetime to be completely foliated by null surfaces. Let us denote any such arbitrary null surface in the family of null surfaces by ${\cal S}$. Further let the (non- affine) parameter  along the null geodesics generating the null surface ${\cal S}$ be denoted by $\lambda$ such that the tangent to the null curves $\el$ is defined to be $\el = \partial / \partial \lambda$. Consider an infinitesimal cross sectional area $\delta A$ of the 2-surface ${\cal S}_t$ which is the intersection of the null surface ${\cal S}$ with the constant time spacelike hypersurface $\Sigma_t$. Then the quantity ${\cal L}_{grav}\delta A d \lambda dt$ is the contribution to the action from a small 4-volume element around the spacetime event on the null surface. Now consider the quantity
\begin{eqnarray}
\frac{{\cal L}_{grav} \delta A }{8\pi} &=& \left[ -\Theta_{ab}\Theta^{ba} - \kappa^2 + (\theta + \kappa)^2 \right]\frac{\delta A }{8\pi} \nonumber \\
&=& - \left( 2 \eta \sigma_{ab}\sigma^{ba} + \zeta \theta^2 \right)\delta A + \frac{\theta \kappa}{4\pi}\delta A
\end{eqnarray}
where we have again set $\Theta^m_n = \sigma^m_n + (1/2) \delta^m_n \theta$ and the viscous co-efficients to be $\eta = 1/16\pi$ and $\zeta = -1/16\pi$. Since $\theta$ gives the fractional rate of change in the null congruence's cross-sectional area $\delta A$, we have
\begin{eqnarray}
(\delta A )\theta = \delta A \left( \frac{1}{\delta A} \frac{d\delta A}{d \lambda}\right) = \frac{d\delta A}{d \lambda}
\label{deltaa}
\end{eqnarray}
Writing $\kappa/4\pi = (\kappa/8\pi) + (1/4)(\kappa/2\pi)$ (the reason for such a splitting is that  pressure is $\kappa/8\pi$ while the temperature is $\kappa/2\pi$; we will say more about this in section \ref{statesection})) and using \eq{deltaa} we get:
\begin{eqnarray}
\frac{{\cal L}_{grav}\delta A}{8\pi} &=& - \delta A \left(2 \eta \sigma_{ab}\sigma^{ba} + \zeta \theta^2 \right) + \frac{\kappa}{2\pi}\frac{d(\delta A /4)}{d\lambda} + \frac{\kappa}{8\pi}\frac{d(\delta A )}{d\lambda} \nonumber \\
&=& - \delta A \left(2 \eta \sigma_{ab}\sigma^{ba} + \zeta \theta^2 \right) + \frac{\kappa}{2\pi}\frac{d(\delta A /4)}{d\lambda} + P\frac{d(\delta A )}{d\lambda}
\end{eqnarray}
where we have used the relation between the pressure and the non-affine parameter $\kappa$ given by $P = \kappa /8\pi$. When the null surface corresponds to a black hole horizon in an asymptotically flat
spacetime, there is a natural choice for $l^a$ such that $\kappa$ can be identified with the surface gravity of the black hole horizon. In a more general context, one can always construct a local Rindler frame around the event on the null surface such that ${\cal S}$ becomes the local Rindler horizon. One can then relate $\kappa$ to the acceleration of the congruence of Rindler observers that has been introduced. Further one can also associate with the horizon a local Unruh temperature $T$ through the relation $T=\kappa /2\pi$ and a Bekenstein entropy $S_{\cal H}$ equal to one-quarter of the cross-sectional area $\delta A$ of the horizon; $S_{\cal H} = \delta A /4$. Then, we obtain
\begin{eqnarray}
\frac{{\cal L}_{grav}\delta A}{8\pi} &=& -\delta A \left(2 \eta \sigma_{ab}\sigma^{ba}+ \zeta \Theta^2 \right) + T\frac{dS_{\cal H}}{d\lambda}+ P\frac{d\delta A}{d\lambda}
\end{eqnarray}
Therefore the temporal rate of change of the action becomes 
\begin{eqnarray}
\frac{1}{\sqrt{-g^\prime}}\frac{d{\cal S}}{dt} &=&  \frac{{\cal L}_{grav}\delta A d\lambda}{8\pi} \nonumber \\
&=& -\delta A d\lambda \left(2 \eta \sigma_{ab}\sigma^{ba}+ \zeta \theta^2 \right) + TdS_{\cal H}+ Pd\delta A \nonumber \\
&=& -dE + TdS_{\cal H}+ Pd\delta A \label{entropyproduction}
\end{eqnarray}
where we have defined $dE \equiv -\delta A d\lambda \left(2 \eta \sigma_{ab}\sigma^{ba}+ \zeta \theta^2 \right)$. The form of \eq{entropyproduction} suggests that we can interpret the rate of change of local action or the entropy density functional as an on-shell local entropy production rate of the given spacetime. Then the three terms in \eq{entropyproduction} can be interpreted as (i) entropy generation due to the loss in energy $dE$ because of viscous dissipation during evolution of the small area element $\delta A$ of the null surface from $\lambda_1$ to $\lambda_2$ along the null congruence (ii) the second term corresponds to rise in the gravitational entropy proportional to the increase of the area of the horizon which is due to the familiar information loss processes (iii) the third term is the (virtual) work done by the horizon against the pressure $P$ during its area expansion $d\delta A$.

\section{Equation of state for the null fluid}\label{statesection}

 We could express the lagrangian in the form in \eq{entropyproduction} by noticing that both temperature and pressure are proportional to the surface gravity $\kappa$ and hence obey the relation $P = T/4$ which is analogous to an equation of state of a gas. If we further use the fact that the entropy per unit area for the horizon is $1/4$, we can write the equation of state as $PA = TS$. Note that in normal units, $P = c^2 \kappa /(8\pi G)$ and has the dimensions of force per unit length as in the case of a two dimensional system of a fluid. In SI units the equation of state can be written as
\begin{equation}
   \underbrace{\frac{c^2 \kappa}{8\pi G}}_{\displaystyle{P}}
    \ \underbrace{A}_{\displaystyle{A}}
  = \underbrace{\frac{\hbar \kappa}{2\pi c k_{B}}}_{
    \displaystyle{k_{B} T}}
 \ \underbrace{\frac{A c^3}{4 \hbar G}}_{
    \displaystyle{S}}
\label{eqnofstate}
\end{equation}
The proportionality between area and horizon entropy used here to write the equation of state $P=T/4$ in the above form, holds in einstein's gravity and is no longer true in the case of higher order curvature theories such as \LL\ theory of gravity. However, it is shown in \cite{membraneLL} that the form of the equation of state holds even in the case of \LL\ theories of gravity apart from an extra constant proportionality factor which purely depends only on the spacetime dimensions $D$ and the order $m$ of the \LL theory considered, and which becomes unity in the case of Einstein's gravity. Further, since all the quantities appearing in \eq{eqnofstate} are defined purely in terms of geometric quantities we can interpret the above equation as describing the equation of state of the underlying microscopic degrees of freedom of emergent gravity. It has been often argued in the literature that it is only the horizons which play a crucial role in exciting or activating these microscopic degrees of freedom. Since the equation of state is valid over the surface of the horizon we find that it is an equation for a system of dimensionality two which makes sense. Further for a two dimensional thermodynamic system, we have the following relation for the Gibbs free energy $G$
\begin{eqnarray}
 G = E - TS + PA
\end{eqnarray}
where $E$ is the energy of the system and the other quantities denote their usual meaning. Here, one can see that an equation of state of the form in \eq{eqnofstate} leads to the conclusion that the Gibbs free energy and the average energy of the system are essentially the same, that is
\begin{eqnarray}
 G = E 
\end{eqnarray}
It is rather intriguing that our hypothetical fluid in the null surface satisfies such an equation of state with $PA=TS$ and it would be nice to see whether one can understand it from any other perspective. Given the fact that black horizons are the very first null surfaces to which the membrane paradigm was applied, we would like to see whether such an equation of state arises dynamically in this context.
We will show, with the help of a gravitating system on the verge of forming a black hole, that the above equation of state does hold for the horizon.

\subsection{Example of a gravitating system}

We consider a system of $n$ densely packed gravitating spherically symmetric shells assumed to be in equilibrium with itself, that is, supporting itself against its own gravity (We closely follow the analysis in \cite{areascaling}). We will show that in the limit when the outermost shell of the system is at a radius very near to its Schwarzschild radius $R = 2M$, where $M$ is the total mass of the system, then the thermodynamic parameters describing the thermodynamic state of outermost shell near the horizon satisfy the same equation of state $P = (1/4)T$.  Let us denote the variables or parameters describing the $i$th shell as $X_i$. Now since the system considered is spherically symmetric, we can write the metric outside the $i$th shell as
\begin{equation}
 ds^2 = -c_i f_i(r_i)dt^2 + (f_i(r_i))^{-1}dr^2 + r_i^2 d\Omega^2
\label{shellmetric}
\end{equation}
where $f_i(r_i)$ is dependent on the mass of the system within a radius $r_i$ and whose functional form is not required for our discussion. Further, the metric has to satisfy the first Israel junction condition which states that the induced metric on a hypersurface should be continuous. This leads to the following constraint on the constants $c_i$:
\begin{equation}
 c_i f_i(r_{i+1}) = c_{i+1} f_{i+1}(r_{i+1})
\label{1junction}
\end{equation}
Now, Birkhoff's Theorem tells us that the metric in the vacuum region outside the $n$th shell should be the Schwarzschild metric, thus we have for the outermost shell $c_n =1$. Then using \eq{1junction}, we can determine the remaining unknown constants to be
\begin{eqnarray}
 c_k = \frac{f_{k+1}(r_{k+1})}{f_{k}(r_{k+1})}.....\frac{f_{n-1}(r_{n-1})}{f_{n-2}(r_{n-1})}\frac{f_{n}(r_{n})}{f_{n-1}(r_{n})}
\label{ccondition}
\end{eqnarray}
Note that when $f_{n}(r_{n}) = 0$ which is true when the outermost shell is exactly on the horizon we have $c_n=0$ for all $(i \neq n)$ which implies that 
\begin{equation}
 g_{00} = 0 \; \; \forall \ i
\label{g00vanish}
\end{equation}
Here the condition that the $g_{00}$'s vanish even for the inner-shells indicates that our assumption regarding the staticity of the inner shells is not valid when $f_{n}(r_{n})$ is exactly zero since we know that a particle cannot be kept at a fixed position inside a blackhole without letting it fall into the singularity. However for the purpose of our discussion, we only need to consider the limit in which the outer shell is very near to the horizon, that is, $f_{n}(r_{n}) \rightarrow 0$, then the statement of $c_i$'s (other than $c_n$) and $g_{00}$ being equal to zero is just the leading order term in this approximation. Henceforth, we shall assume that we are working in this limit and will not state it explicitly unless otherwise needed.\\ \\
The second junction condition gives us the surface stress energy tensor $T_{\alpha \beta}$ on the shell which can be determined by the following equation
\begin{equation}
 8\pi (T^{\alpha}_{\beta})_i = \left[ \delta^{\alpha}_{\beta} K - K^{\alpha}_{\beta} \right]_i
\end{equation}
Here $K^{\alpha}_{\beta}$ is the extrinsic curvature of the shell and $[ \ ]_i$ denotes the jump in the quantities, that is, $[h(r)]_i = h(r_{i}) - h(r_{i-1})$. We can determine the energy $E$ and pressure $P$ of the $i$th shell by using the static nature of the shells and define the energy as $E_i = -4\pi r_i^2 T^0_0$ while the pressure (tangential to the surface) is defined to be $P_i = T^{\theta}_{\theta}$. The physical meaning associated with $E$ and $P$ is same as the energy and pressure measured by a local observer at rest on the shell. Using the form of the metric in \eq{shellmetric}, we find 
\begin{eqnarray}
 E_i &=& - r_i \left[ \sqrt{f} \right]_i \label{Eform} \\
8\pi P_i &=&  \frac{1}{2} \left[ \frac{f^{\prime}}{\sqrt{f}} \right]_i + \frac{1}{r_i} \left[ \sqrt{f} \right]_i
\label{Pform}
\end{eqnarray}
The other thermodynamic parameters such as $T$ and $\mu$ can be found from the condition of thermodynamic equilibrium as follows. Thermal equilibrium implies that the temperature $T_i$ obeys the Tolman relation
\begin{eqnarray}
 T_i &=& \frac{T}{\sqrt{-g^i_{00}(r_i)}}
\label{tolman}
\end{eqnarray}
where $T$ is the temperature of the system as measured by a static observer at infinity, whereas the condition for chemical equilibrium implies the chemical potential $\mu_i$ to satisfy
\begin{eqnarray}
\mu_i \sqrt{-g^i_{00}(r_i)} &=& \mu_n \sqrt{-g^n_{00}(r_n)}
\label{chemical}
\end{eqnarray}
Further, in thermodynamic equilibrium, each shell satisfies the Gibb's Duhem relation
\begin{equation}
 E_i = T_i S_i - A_i P_i + \mu_i N_i
\end{equation}
where $N_i$ is the number of particles composing the $i$th shell. Using \eq{tolman} and \eq{chemical} in the above expression, we can solve for the total entropy $S$ of the system to write it in the form
\begin{equation}
 S = \sum_i \frac{E_i + P_i A_i}{T}\sqrt{-g^i_{00}(r_i)} - \frac{\mu_n N}{T}\sqrt{-g^n_{00}(r_n)}
\end{equation}
Now assuming that $\mu_n N$ is a finite quantity, the last term in the above expression vanishes in the near horizon limit, since $g_{00}$ vanishes (see \eq{g00vanish}). (In the case of the system comprising of only photons, the last term is zero since $\mu =0$ for photons). Hence the only non-zero contribution to the entropy can come from the first term provided the prefactor of $\sqrt{g_{00}}$ contains a divergent term of the order $(1/\sqrt{g_{00}})$ or higher. Now from the expressions of $E_i$ and $P_i$ in \eq{Eform} and \eq{Pform}, one can check that it is only the first term $\sim f^{\prime}/\sqrt{f}$ in the expression of $P_n$ of the required order that makes a non-zero contribution to the total entropy. Hence, we can write $P_n$ to the same leading divergent order as
\begin{eqnarray}
P_n &\approx & \frac{1}{16\pi}\frac{f_n^{\prime}}{\sqrt{f_n}} = \frac{1}{8\pi}\frac{\kappa}{\sqrt{f_n}} \nonumber \\
&=& \frac{1}{4} \frac{T_H}{\sqrt{f_n}}
\label{pressure}
\end{eqnarray}
where $\kappa$ is the surface gravity of the horizon and $T_H = \kappa/(2\pi)$ is the Hawking temperature of the horizon. Now, if we assume that the outermost shell which is at rest very near to horizon has come to be in thermal equilibrium with the horizon temperature, then we have $T_n = T_H/\sqrt{f_n}$ and hence
\begin{eqnarray}
P_n \approx \frac{1}{4} T_n
\label{shellstate}
\end{eqnarray}
Thus we find that the outermost shell near to the horizon satisfies the same equation of state as in the case of the DNS equation. Further, the total entropy $S$ gets a non-zero contribution only from the tangential pressure of the outermost shell and we have
\begin{eqnarray}
 S &\approx& S_n \approx \frac{P_n A_n}{T_n} \\
&\approx& \frac{1}{4}A_n
\end{eqnarray}
which is same as the Bekenstein-Hawking entropy for a black hole (The area scaling of entropy for a gravitating system has been discussed before in \cite{areascaling}). One can note, in this case that the origin of the $(1/4)$ factor in the expression of entropy is due to the equation of state of \eq{shellstate}.

\section{Conclusions}\label{sect3}

The emergent paradigm of gravity is based on the idea that the usual field equations of gravity arise in the long wavelength limit when we average over suitable microscopic degrees of the --- as yet unknown --- underlying theory of quantum gravity. In the absence of the such a microscopic theory, one can at present only demonstrate the possible emergent behavior by comparing gravity with other emergent physical processes known in nature such as thermodynamics, fluid mechanics etc. In this context, obtaining  a set of equations very similar to those of fluid mechanics directly from a thermodynamic extremum principle is an important step, which has been achieved in this paper.

The nature of extremum principle and the structure of the corresponding Lagrangian should contain possible information about the manner in which gravity becomes emergent. In this context, we note that the Lagrangian density of the gravitational part can be written in several algebraically equivalent forms, each of which deserves further exploration:
\begin{eqnarray}
-{\cal L} &=& 2 P^{abcd} \nabla_c l_a \nabla_d l_b  \label{paddyaseemm} \\
&=&-\Theta_{ab}\Theta^{ba} - \kappa^2 + (\theta + \kappa)^2 \label{actionfluid} \\ 
&=& -8\pi \left( 2 \eta \sigma_{ab}\sigma^{ba} + \zeta \theta^2 \right) + 2\theta \kappa \label{actionfluid1} \\
&=& -\chi_{ab}\chi^{ba} + \chi^2 \label{actionfluid2} 
\end{eqnarray}
The form in \eq{paddyaseemm} was used earlier and explored extensively in \cite{aseemtp}. If we think of spacetime as analogous to an elastic solid, then the diffeomorphism $x^a \rightarrow x^a + \xi^a$ can be thought as as analogous to the elastic deformation of the solid. Such a distortion, in general, is not of much relevance to our consideration except when it deforms the \textit{null surfaces} of the spacetime. If we consider a small patch of  null surface  as a part of a local Rindler horizon of suitable class of observers in the spacetime, the deformation of the null surface changes the accessibility of information by these observers.  Given the intimate connection between information and entropy, this leads to the variation of entropy as measured by these observers due to the deformation. In other words, it seems reasonable to assume that deforming a null surface should  cost  entropy. Taking a cue from the usual description of macroscopic solids, elasticity etc., we would expect the leading term in the entropy change to be a quadratic functional of the displacement field $\xi^a$ which is precisely what we have in \eq{paddyaseemm}.

The analogy between gravity and a viscous fluid is further strengthened by the form of the lagrangian in Eq.(\ref{actionfluid}, \ref{actionfluid1},  \ref{actionfluid2}). We have shown that starting from this lagrangian one can directly derive the \nsle (i.e., Eq.(\ref{rmneqn})) without the need of first deriving the Einstein field equations and then projecting it suitably on the horizon. On the other hand, we know that the Einstein field equations also follow from the same action expressed in the form of \eq{paddyaseemm}, thereby showing the equivalence between the two interpretations. In extremizing the functional in  \eq{paddyaseemm}, we demand that the extremum condition holds for all null vectors in the spacetime, which is equivalent to demanding the validity of the extremum principle for all local Rindler observers in the spacetime. While using the form of the functional in \eq{actionfluid}, say, we do something similar in the sense that we demand the resulting DNS equations hold for all null surfaces. But now we express the result in terms of fluid variables which, in turn, are defined in terms of the null vector itself.
It is rather curious that such an interpretation leads to an equation of state of the form $PA = TS$, the physical meaning of which is at present unclear. It is, however, interesting to note that any microscopic description should eventually lead to a long wavelength limit in which this equation of state emerges in a natural form. 

It may also be noted that the lagrangian density in \eq{paddyaseemm} obeys the relation:
\begin{equation}
 \frac{\partial {\cal L}}{\partial ( \nabla_c \ell^a)}  \propto (\nabla_a \ell^c  - \delta^c_a \nabla_i \ell^i)
\label{brownyorktensor}
\end{equation} 
This term is analogous to the more familiar Brown-York tensor $t^c_a = K^c_a - \delta^c_a K$, where $K_{ab}$ is the extrinsic curvature that arises in the (1+3) separation of Einstein's equations. (More precisely, the appropriate projection to 3-space leads to $t^c_a$.) This combination can be interpreted as a surface energy momentum tensor in the context of membrane paradigm because $t_{ab}$ couples to $\delta h^{ab}$ on the boundary surface when we vary the gravitational action. In fact, one obtains the results for null surfaces as a limiting process from the time-like surfaces (usually called stretched horizon) in the case of membrane paradigm \cite{membrane}. Equation(\ref{brownyorktensor}) shows that the entropy functional is related to  $t^c_a$ and its counterpart in the case of null surface. One may also note that starting from the Brown-York tensor, the gravity fluid/duality for Rindler spacetimes was demonstrated in \cite{rindlergravityfluidduality}.

The bulk viscosity of the black hole obtained through the membrane paradigm comes out to be negative and has been subject of debate in the recent gravity/fluid duality studies since the bulk viscosity of a real fluid must be positive. In proper hydrodynamics, this identification leads to a local entropy decrease. It has been argued  that for hydrodynamic concepts such as transport coefficients to have physical sense, there must be a separation of scales between the temperature (say) and the hydrodynamic gradients. In the original membrane paradigm, as opposed to the modified version of fluid/gravity duality inspired by AdS/CFT, there is no such separation of scales. This is because the stress-energy tensor of the membrane is conserved through a covariant derivative in a metric whose curvature (e.g. surface gravity) is of the same order of magnitude as the temperature. In contrast, in AdS/CFT-like fluid/gravity duality, the two required scales are well separated and one can then define a small parameter which can be used in the derivative expansion of the hydrodynamic gradients to determine them upto different orders. The bulk viscosity for pure Einstein gravity in this case is determined to be zero (while the the bulk viscosity of a real non-conformal fluid must be positive , see \cite{C.eling}). In view of this, the connection between gravity and a viscous fluid through the membrane paradigm should be thought of as somewhat formal arising mainly through the analogy between the stress-energy tensor of the membrane and the hydrodynamic stress-energy tensor and hence, in our opinion, deserves attention provided one is consistent in the concepts within its framework. Moreover, one of the main aims of this paper was to show the emergent nature of the action which leads to the gravity/fluid duality. This we have achieved by demonstrating that the action can be interpreted as a local entropy production rate by consistently using the concepts of the membrane paradigm and working strictly within its own domain. Further, we believe that the techniques used in the AdS/CFT-like fluid/gravity duality can be extended to any null surface and will be dealt in a separate future publication (in progress).

\section*{Acknowledgements}
We thank D. Kothawala for comments on an earlier draft. SK is supported by a Fellowship from the Council of Scientific and Industrial Research (CSIR), India and the research of TP is partially supported by the J.C. Bose fellowship of DST, India. 

\appendix{}
\section*{Appendix: The momentum density $\Omega_a$}\label{momentum}

In section \ref{sect1}, we recalled that the momentum density $-\Omega_a/8\pi$ could be determined by comparing the projected Einstein's equation Eq.[\ref{rmneqn}] with the form of the Navier-Stokes equation for a viscous fluid. It was shown to depend on the geometric quantities defining the null surface through the relation in \eq{defOmega}, that is:
\begin{equation}
\Omega_b \equiv q^a_b \omega_a    = - q^a_b (k_m \ch ma ) = \omega_b  - \kappa k_b ( k_m \ell^m) = \omega_b + \kappa k_b                                                                                                                                                                                                                                                                                                                                                                                                                                                                                                             \end{equation}
Apart from the analogy of the projected Einstein's equation with Navier-Stokes equation for a viscous fluid, any further motivation for calling the above geometric quantity as momentum density seems to be  missing in literature. It is known only in the case of a Kerr black hole spacetime, that the momentum density $-\Omega_a/8\pi$ when integrated over the two dimensional horizon surface is equal to the total angular momentum $J$ of the rotating black hole, which of course was a crucial part in the study of the membrane paradigm for Black holes. In contrast, the shear tensor $\sigma^m_{a}$ when expressed in a co-ordinate system suitably adapted for describing the null surface (see \eq{adapted}) has a form
\begin{equation}
\Theta_{AB}=\frac{1}{2}\left( D_Av_B+D_Bv_A+ \frac{\partial q_{AB}}{\partial t}\right)
\label{imp1} 
\end{equation} 
which reduces  (when $q_{AB}$ is independent of $t$) to that of the shear tensor of a viscous fluid with velocity field $v_A$. This is the key reason why the projected equations can be interpreted as the Navier-Stokes equation. It would be interesting if one could also express the geometrical definition of the momentum density $-\Omega_A/8\pi$ in a suitable form which could make its physical meaning apparent, in particular if it could be shown to be proportional to $v_A$.  

In this appendix, we explore the form of the momentum density $-\Omega_a/8\pi$ by expressing it in the following adapted co-ordinate system.
\begin{equation}
ds^2 = - N^2 dt^2 + \left(\frac{M}{N} dx^3+\epsilon Ndt \right)^2 + q_{AB} (dx^A- v^ A dt + m^A dx^3) ( dx^B- v^ B dt + m^B dx^3) \\
\label{adapted}
\end{equation} 
Here, $x^3$ = constant defines the horizon $S$ while the metric on $S_t$ corresponding to $t$ = constant, $x^3$ = constant is $q_{AB}$. $x^A$ are the co-ordinates covering the surface $S_t$.
In the adapted co-ordinate system, the momentum density $\Omega_A $ is just $ \omega_A$ which we can express as
\begin{eqnarray}
w_A &=& -k^m \Pi^n_A \nabla_n l_m =^* l^m \nabla_m k_A = \Gamma^0_{Am}l^m \nonumber \\
&=& - \frac{v^B}{2M}\partial_1 q_{AB} - \frac{m^B}{2M}\partial_0 q_{AB} - \left[ \frac{m^B}{2M}\left( D_B v_A - D_A v_B \right)+ \frac{v^B}{2M}\left( D_B m_A + D_A m_B \right) \right] 
\end{eqnarray}
The first term can be thought of as describing the rate of flow of $q_{AB}$ along the null direction. The second term vanishes when we demand that $\partial q_{AB}/\partial t = 0$ in accordance with \eq{imp1} for $\sigma^m_{a}$ to be interpreted as the shear tensor. The third term measures the momentum in the curl of the velocity field $v_A$ while the fourth term shows that we can associate an inertia ${\cal M}_{AB} = (D_B m_A + D_A m_B)/2M$ such that $v^B{\cal M}_{AB}$ is the momentum density associated with the flow. Since the metric coefficient (time-2 space) $g_{0A} = -v_A$ behaves as a velocity field, we can in the same way associate a flow with the co-efficient (null-2 space) $g_{1A} = m_A$, then the stresses in this flow ${\cal M}_{AB} = (D_B m_A + D_A m_B)/2M$ provides an inertia for the velocity field $v_A$.

Consider now  the case when $m^A =0$, when the expression for the momentum density reduces to just one term
\begin{eqnarray}
w_A &=& - \frac{v^B}{2M}\partial_1 q_{AB}
\end{eqnarray}
Since $q_{AB}$ is the metric of the two-dimensional surface $S_t$, it is always possible to diagonalize it, that is, we can write $q_{AB} = q_{22}\delta_{A2}\delta_{B2}+q_{33}\delta_{A3}\delta_{B3}$ for the two dimensional surface. Hence, we have 
\begin{eqnarray}
w_A &=& -\frac{v_A}{2M}\frac{\partial_1 q_{AA}}{q_{AA}} \; \; \; \mathrm{(no\ summation)}
\label{momdenshort}
\end{eqnarray}
$w_A$ is proportional to $v_A$. We can now read off the  the inertia by comparing the above expression with $p = mv$ for a non-relativistic fluid momentum density. We see that $\partial_1 q_{AA}/(2Mq_{AA})$ then plays the role of mass density for the flow. Note that to obtain the momentum density in the form of \eq{momdenshort}, the only extra condition we have to assume is $m^A =0$  which can always be imposed by a suitable choice of coordinates \cite{paddynavier}. However, this may not be the most natural coordinate system for the study of the problem in question and one often has to work with coordinate systems in which $m^A \neq 0$. 
It would be interesting to determine the physical meaning of the extra terms which arise when $m^A \neq 0$, which we hope to address in a future publication.

\end{document}